\documentclass[twocolumn]{aastex631}

\newcommand{\mpc}{\mathrm{~km~s^{-1}~Mpc^{-1}}}

\usepackage{subfigure}
\usepackage{amsmath}
\usepackage{hyperref}
\usepackage{multirow}
\usepackage{bm}

\shorttitle{Measurements of $H_0$ and $\Omega_{K}$ by SGL and GW}
\shortauthors{Meng-Di Cao et al.}

\begin{document}

\title{A new way to explore cosmological tensions using gravitational waves and strong gravitational lensing}

\author{Meng-Di Cao}
\affiliation{Department of Astronomy, Beijing Normal University, Beijing 100875, China}

\author{Jie Zheng}
\affiliation{Department of Astronomy, Beijing Normal University, Beijing 100875, China}

\author{Jing-Zhao Qi}
\affiliation{Department of Physics, College of Sciences, \& MOE Key Laboratory of Data Analytics and Optimization for Smart Industry, Northeastern University, Shenyang 110819, China}

\author{Xin Zhang}
\affiliation{Department of Physics, College of Sciences, \& MOE Key Laboratory of Data Analytics and Optimization for Smart Industry, Northeastern University, Shenyang 110819, China}

\author{Zong-Hong Zhu}
\affiliation{Department of Astronomy, Beijing Normal University, Beijing 100875, China}

\correspondingauthor{Jing-Zhao Qi} \email{qijingzhao@mail.neu.edu.cn}

\begin{abstract}
In recent years, a crisis in the standard cosmology has been caused by inconsistencies in the measurements of some key cosmological parameters, Hubble constant $H_0$ and cosmic curvature parameter $\Omega_K$ for example. It is necessary to remeasure them with the cosmological model-independent methods. In this paper, based on the distance sum rule, we present such a way to constrain $H_0$ and $\Omega_K$ simultaneously in the late universe from strong gravitational lensing time delay (SGLTD) data and gravitational wave (GW) standard siren data simulated from the future observation of the Einstein Telescope (ET). Based on the currently 6 observed SGLTD data, we find that the constraint precision of $H_0$ from the combined 100 GW events can be comparable with the measurement from SH0ES collaboration. As the number of GW events increases to 700, the constraint precision of $H_0$ will exceed that of the \textit{Planck} 2018 results. Considering 1000 GW events as the conservative estimation of ET in ten-year observation, we obtain $H_0=73.69\pm 0.36 \mpc$ with a 0.5\% uncertainty and $\Omega_K=0.076^{+0.068}_{-0.087}$. In addition, we simulate 55 SGL systems with 6.6\% uncertainty for the measurement of time-delay distance. By combining with 1000 GWs, we infer that $H_0=73.65\pm0.35 \mpc$ and $\Omega_K=0.008\pm0.048$. Our results suggest that this approach can play an important role in exploring cosmological tensions.

\end{abstract}


\section{Introduction} \label{sec:intro}

It is well known that the most serious crisis in modern cosmology is the Hubble tension problem. Specifically, as a fundamental cosmological parameter representing the expansion rate of the present universe, the Hubble constant $H_0$ could be constrained by global fitting from cosmological observations. In this way, measurements of temperature and polarization anisotropies in the cosmic microwave background (CMB) of the early universe from \textit{Planck} satellite predict a value of $H_0 = 67.4 \pm 0.5 \mpc$ \citep{Planck:2018vyg} in the framework of the flat $\Lambda$ cold dark matter ($\Lambda$CDM) model. Alternatively, the local Type Ia supernovae (SN Ia) calibrated by the distance ladder could be used to estimate $H_0$ through Hubble-Lema\^{\i}tre law without assuming any cosmological model, yielding $H_0 = 74.03 \pm{1.42} \mpc$ from SH0ES (SNe, H0, for the Equation of State of dark energy) collaboration \citep{Riess:2019cxk}. There is a $4.4\sigma$ tension between the two independent estimates of $H_0$, which cannot be simply attributed to systematic error \citep{DiValentino:2018zjj,Feeney:2017sgx,Follin:2017ljs,Riess:2019cxk}. This crisis reveals that there could be an inconsistency between the early universe and the late universe \citep{Verde:2019ivm}.

On the other hand, recent studies \citep{Planck:2018vyg,DiValentino:2019qzk,Handley:2019tkm} concerning the cosmic curvature parameter $\Omega_K$ presented a new challenge to cosmology. The value of curvature parameter $\Omega_K$ determines the shape of the universe with $\Omega_K = 0$ representing a spatially flat universe, whereas $\Omega_K < 0$ and $\Omega_K > 0$ correspond to a spatially closed and open universe, respectively. The anomalous lensing amplitude, $A_{\rm{lens}} > 1$, inferred by the CMB temperature and polarization data from \textit{Planck} suggests a closed universe ($\Omega_K < 0$) at $99\%$ confidence level \citep{Planck:2018vyg}. However, the combination of CMB data and baryon acoustic oscillation (BAO) measurements point toward a spatially flat universe with a remarkable $0.2\%$ precision, which means there is an inconsistency between \textit{Planck} and BAO data for the constraint on the curvature parameter. This is even directly called "curvature tension" by \citet{Handley:2019tkm}. Moreover, a closed universe as \textit{Planck} prefers will exacerbate the Hubble tension and $S_8$ tension \citep{DiValentino:2019qzk,DiValentino:2020hov}, which makes it possible for a larger discordance to be hidden in a flat $\Lambda$CDM model.

All of these crises implies something is not well understood, and there may well be new physics at play. Anyway, many of the conclusions we have known for certain now need to be reconfirmed. In particular, it is necessary to measure fundamental cosmological parameters in the late universe with the new and cosmological model-independent ways. Recently, based on the distance sum rule, \citet{Collett:2019hrr} proposed a method with combining the observations of strong gravitational lensing time delay (SGLTD) and type Ia supernova (SN Ia) luminosity distance to determine $H_0$ and $\Omega_K$ simultaneously without dependence on any cosmological model. In this scheme, the SN Ia as the standard candle is used to calibrate the three distances of the strong gravitational lensing (from observer to lens, from observer to source, and from lens to source). However, it is important to note that uncalibrated SN Ia cannot provide absolute distance but relative distance. The absolute luminosity distance can only be obtained if the absolute magnitude $M_B$ of SN Ia is determined. Due to $M_B$ being exactly degenerate with $H_0$, using SN Ia to determine $H_0$ with this method is theoretically problematic. After that, \citet{Wei:2020suh} extended this approach by using the known ultraviolet versus X-ray luminosity correlation of quasars to calibrate the distances of SGLTD and determined $H_0$ and $\Omega_K$. \citet{Qi:2020rmm} combined 7 SGLTD observations and 120 intermediate-luminosity quasars calibrated as standard rulers and obtained stringent constraints on $H_0$ and $\Omega_K$. More recently, with the same approach, \citet{Qi:2022sxm} made a forecast of using the strongly lensed SN Ia to improve the measurements on $H_0$ and $\Omega_K$.

Recently, with the successful detections of gravitational waves (GWs) by LIGO and VIRGO detectors \citep{LIGOScientific:2016aoc,LIGOScientific:2016sjg,LIGOScientific:2017bnn}, the era of GW astronomy and multi-message astronomy is coming. Compared to traditional cosmological probes, a great advantage of GW is that the standard siren could provide the absolute luminosity distance \citep{Schutz:1986gp,LIGOScientific:2017adf} without any calibration, which can play an extremely important role in cosmological studies \citep{Qi:2021iic,Pan:2021tpk,Zhao:2010sz,Wang:2018lun,Zhang:2019ylr,Wang:2019tto,Zhang:2019loq,Zhao:2019gyk,Jin:2020hmc,Wang:2021srv,Jin:2021pcv}. Up to now, the available standard siren data are too few to make a significant contribution to cosmology. However, the next-generation ground-based GW observatory, such as Einstein Telescope (ET) with 10 km-long arms and three detectors, is expected to detect about 1000 standard sirens from the events of binary neutron star (BNS) merger in ten-year observation \citep{Nissanke:2009kt,Zhao:2010sz,Cai:2016sby,Zhang:2020axa}. This remains us to explore what role GW will play in the measurements of $H_0$ and $\Omega_K$ with the cosmological model-independent way in the era of GW astronomy. Actually, GW standard siren measurement can determine $H_0$ directly at low redshift through Hubble-Lema\^{\i}tre law, and that has been implemented to the GW170817 event with a poor precision \citep{LIGOScientific:2017adf}. However, Hubble-Lema\^{\i}tre law is valid only in the low-redshift region $z < 0.1$. The method concerned in this paper is valid in a higher range of redshift, which allows us to explore the Hubble tension problem at high redshift, but also to investigate the cosmic curvature problem in the late universe simultaneously.

In this paper, we will investigate the constraints on the Hubble constant and the cosmic curvature parameter with SGLTD and GW from two aspects. Firstly, we simulate 1000 GW events data in the redshift range of $0 < z \leq 2$ based on ET in ten-year observation, and then constrain $H_0$ and $\Omega_K$ in combination with 6 observed SGLTD data. On the other hand, the ET project foresees the beginning of construction in 2026 with the goal to start observations in 2035. By then, future surveys, like the Large Synoptic Survey Telescope (LSST) with wide field-of-view and frequent time sampling to monitor the SGL systems for time delay measurements, will provide a large sample of well-measured SGLTD observations. Therefore, we will also simulate SGLTD data based on LSST and combine GW data from ET to predict what precision $H_0$ and $\Omega_K$ can be constrained by this method in the future.

\section{Methodology and data} \label{sec:method}

Assuming our universe is homogeneous and isotropic, the spacetime can be described by the Friedmann-Lema\^{\i}tre-Robertson-Walker metric
\begin{equation}
d s^{2}=c^{2} d t^{2}-\frac{a(t)^{2}}{1-K r^{2}} d r^{2}-a(t)^{2} r^{2} d \Omega^{2} ,
\label{Eq:Eq1}
\end{equation}
where $c$ is the speed of light and $a(t)$ is the scale factor. $K$ is a constant representing the spacial curvature, which is related to the cosmic curvature parameter $\Omega_K$ as $\Omega_{K}=-K c^{2} / a_{0}^{2} H_{0}^{2}$. The Hubble parameter is defined as $H \equiv \dot{a} / a$, and $H_0$, Hubble constant, represents the present value of $H(z)$. For a SGL system, the angular diameter distance between the lens at redshift $z_l$ and the source at redshift $z_s$ can be denoted as $D_{A}\left(z_{l},z_{s}\right)$. The dimensionless comoving distance $d\left(z_{l}, z_{s}\right) \equiv\left(1+z_{s}\right) H_{0} D_{A}\left(z_{l}, z_{s}\right)/c$ could be given by
\begin{equation}
d\left(z_{l}, z_{s}\right)=\frac{1}{\sqrt{\Omega_{K}}} \operatorname{sinn} \left(\sqrt{\Omega_{K}} \int_{z_{l}}^{z_{s}} \frac{H_{0}}{H(z)} d z\right),
\label{Eq:Eq2}
\end{equation}
where
\begin{equation}
\operatorname{sinn}(x)= \begin{cases}\sin (x), & \Omega_{K}<0, \\ x, & \Omega_{K}=0, \\ \sinh(x) , & \Omega_{K}>0.\end{cases}
\end{equation}
By introducing $d_{ls} \equiv d\left(z_l, z_s\right)$, $d_{l} \equiv d\left(0,z_{l}\right)$ and $d_{s} \equiv d\left(0,z_{s}\right)$, these three distances are connected by the well-known distance sum rule \citep{Rasanen:2014mca,Xia:2016dgk,Li:2018hyr,Liao:2019hfl,Qi:2018atg,Qi:2018aio,Wang:2020dbt,Zhou:2019vou,Wang:2019yob,Wang:2022rvf}:
\begin{equation}
\frac{d_{l s}}{d_{s}}=\sqrt{1+\Omega_{K} d_{l}^{2}}-\frac{d_{l}}{d_{s}} \sqrt{1+\Omega_{K} d_{s}^{2}}.
\label{Eq:Eq3}
\end{equation}
Furthermore, Equation (\ref{Eq:Eq3}) can be rewritten as
\begin{equation}
\frac{d_{l} d_{s}}{d_{l s}}=\frac{1}{\sqrt{1 / d_{l}^{2}+\Omega_{K}}-\sqrt{1 / d_{s}^{2}+\Omega_{K}}}.
\label{Eq:Eq4}
\end{equation}
The left-hand side of Equation (\ref{Eq:Eq4}) can be obtained from the measurements of the SGL time delay. If the distances in the right-hand side of Equation (\ref{Eq:Eq4}) are determined, $H_0$ and $\Omega_K$ involved in the distance sum rule can be constrained without dependence on any cosmological model.

\subsection{GW simulation}
To constrain $H_0$ and $\Omega_K$ in the framework of distance sum rule requires to determine the distances $d_{l}$ and $d_{s}$ \citep{Collett:2019hrr,Cao:2019kgn}. In this work, we use gravitational-wave signals from the binary neutron star (BNS) mergers, known as ``standard sirens'', to calibrate $d_{l}$ and $d_{s}$. Since the number of GW standard sirens observed so far is not enough for distance calibration, we choose to use simulated GW data for distance analysis. Following the process in previous works \citep{Zhao:2010sz,Sathyaprakash:2009xt,Cai:2016sby}, we generate 1000 GW events based on the third-generation ground-based GW detector ET in ten-year observation. For the detectable redshift range of electromagnetic counterpart, above previous studies considered that the upper redshift for ET is $\sim$ 5. However, by investigating various models of the short $\gamma-ray$ bursts and afterglows, a recent study concluded that the redshift limit of detectable electromagnetic counterpart is $z \leq 2$ for the third-generation ground-based GW detector \citep{Yu:2021nvx,Chen:2020zoq}. Therefore, we conservatively adopt the redshift range of $0 < z \leq 2$ for the simulation of GW in this work. The redshift distribution of the BNS sources takes the form \citep{li2015extracting}
\begin{equation}\label{p}
P(z) \propto \frac{4 \pi d_{C}^2(z) R(z)}{H(z)(1+z)}, 
\end{equation}
where $d_{C}(z)$ represents the comoving distance at the redshift $z$. $R(z)$ is the burst-rate function per unit source time and unit comoving volume. According to the population synthesis models and the cosmic star formation history \citep{Schneider:2000sg}, \citet{Cutler:2009qv} created a piece-wise fit for $R(z)$,
\begin{equation}\label{R}
R(z)=  \begin{cases}
  z+2z, & \text{$z \le 1$},\\
  \frac{3}{4}(5-z), & \text{$1 < z < 5$},\\
  0, & \text{$z \ge 5$}.
  \end{cases}
\end{equation}

The simulation method in this paper follows the prescription in previous works \citep{Zhao:2010sz,Sathyaprakash:2009xt,Cai:2016sby,Qi:2019wwb}, and then we do not repeat it here. It is worth pointing out that a gravitational waveform is determined by some specific parameters, such as the mass ratio of BNS, the position angle of the source, and so on. The random sampling of parameters in the simulation process usually leads to a random bias. To eliminate this bias, we perform $6\times10^4$ Monte Carlo simulations and select the most probable values of luminosity distances and uncertainties.

For the standard siren, the analysis of GW’s waveform could give the absolute luminosity distance $D_{L}$. In simulation, the fiducial model we choose is the $\Lambda$CDM model with $\Omega_{m} = 0.315$ and $H_{0} = 74.03 \mpc$, where $\Omega_{m}$ is the current matter density in units of the critical density and its value adopted is from \textit{Plank} 2018 results \citep{Planck:2018vyg}. The value of $H_{0}$ we adopt is taken from the local measurements by SH0ES \citep{Riess:2019cxk} because we are investigating the late universe in this paper.

Following the error strategy described in \cite{Zhao:2010sz,Cai:2016sby}, the total uncertainty on luminosity distance depends on the instrumental error $\sigma_{D_{L}}^{\text {inst}}$ and an additional error $\sigma_{D_{L}}^{\text {lens}}$ caused by the weak lensing, and its expression is

\begin{equation}
\sigma_{D_{L}}=\sqrt{\left(\sigma_{D_{L}}^{\text {inst }}\right)^{2}+\left(\sigma_{D_{L}}^{\text {lens }}\right)^{2}}
\label{Eq:Eq5}
\end{equation}
The instrumental error of the luminosity distance is related to the signal-to-noise ratio (SNR) $\rho$ as $\sigma_{D_{L}}^{\text {inst }} \simeq D_{L}/\rho$ \citep{li2015extracting}. However, the luminosity distance $D_L$ is correlated with other GW parameters, especially the inclination angle $\iota$ of the binary's orbital angular momentum with the line of sight. The maximal effect of the inclination angle on the SNR between the source being face-on ($\iota=0$) and edge-on ($\iota=\pi/2$) is estimated as a factor of 2 \citep{li2015extracting}. Therefore, to consider the correlation between the $D_L$ and $\iota$, we estimate the instrumental error as $\sigma_{D_{L}}^{\text {inst }} \simeq 2 D_{L}/\rho$. Besides, the lensing uncertainty caused by the weak lensing is modelled as $\sigma_{D_{L}}^{\text {lens }}=0.05 z D_{L}$ \citep{Sathyaprakash:2009xt}.

By using GW data to calibrate the distances of the SGL system, the difficulty is that the redshifts of SGL data and GW data cannot be one-to-one correspondence exactly. The previous way of treating this issue is by reconstructing a continuous distance function using a polynomial fit \citep{Qi:2020rmm,Collett:2019hrr,Wei:2020suh}. Here, we also reconstruct $d(z)$ by fitting the luminosity distances from 1000 simulated GW data with a third-order polynomial,
\begin{equation}
d(z)=z+a_{1} z^{2}+a_{2} z^{3},
\label{Eq:Eq6}
\end{equation}
and with the initial conditions of $d(0)=0$ and $d'(0)=1$. For convenience, firstly we convert the luminosity distance of GW into the dimensionless distance via $d(z)=H_{0} D_{L}(z) /(c(1+z))$. Now distances $d_{l}$ and $d_{s}$ corresponding to the lens redshift and source redshift could be easily obtained from Equation (\ref{Eq:Eq6}).

For simulated GW data, the likelihood estimator $\chi_{\rm{GW}}^{2}$ is given by
\begin{equation}
\chi_{\rm{GW}}^{2}=\sum_{i=1}^{1000}\left(\frac{D_{L}^{\rm{GW}}\left(z_{i}\right)-d(z_i)(1+z_i)\frac{c}{H_0}}{\sigma_{D_{L, i}}}\right)^{2}.
\label{Eq:Eq7}
\end{equation}
Then we could calculate the likelihood $\mathcal{L}_{\mathrm{GW}}$ from the likelihood estimator, $\mathcal{L}_{\mathrm{GW}} \sim \exp \left(-\chi_{\mathrm{GW}}^{2} / 2\right)$. 

To check the validity of our approach based on the third-order polynomial expansion, for comparison, we use third-order, fourth-order, and fifth-order polynomials to fit the simulated GW data by maximizing the likelihood function $\mathcal{L}_{\mathrm{GW}}$. To quality the goodness of fitting with different order polynomials, we use Bayesian information criterion (BIC) as an evaluation tool \citep{Schwarz:1978tpv}, and its expression is 
\begin{equation}
\mathrm{BIC}=-2 \ln \mathcal{L}_{\max }+k \ln N,
\end{equation}
where $\mathcal{L}_{\max }$ is the maximum likelihood, $k$ is the number of model parameters, and $N$ is the number of data points. In general, a model with a smaller value of BIC is more favored by observations. $\Delta\mathrm{BIC}=2$ is considered to be positive evidence, and $\Delta\mathrm{BIC}\leq 6$ is regarded as strong evidence against the model with higher BIC value \citep{Mukherjee:1998wp,Liddle:2004nh}.

In Table \ref{table:bic}, we list the fitting results of 3rd, 4th and 5th-order polynomials. It can be seen that the BIC value of the 3rd-order polynomial is much smaller than that of the other two polynomials, which strongly indicates that the 3rd-order polynomial is flexible enough for the distance reconstruction. Therefore, we will use the 3rd-order polynomial to reconstruct $d(z)$ from 1000 simulated GW data in this paper.

\begin{table*}[htbp]
\begin{center}{
\caption{Constraint results of 3rd, 4th and 5th-order polynomials from the simulated GW data, and corresponding BIC values.  \label{table:bic}}
\begin{tabular}{c|ccccc} \hline\hline
The Order of Polynomial &$a_1$  &$a_2$ &$a_3$ &$a_4$ & BIC  \\
  \hline
3rd & $-0.274\pm 0.006$&  $0.036\pm 0.003$& - &- & $14.03$ \\
4th & $-0.270\pm 0.015$& $0.030\pm 0.022$&  $0.002\pm 0.008$& - & $20.87$\\
5th &$-0.271\pm 0.028$&  $ 0.030\pm 0.080$&  $0.002\pm 0.070$&  $0.000\pm 0.019$ & $27.78$\\
\hline
\hline
 \end{tabular}
 }
\end{center}
\end{table*}

\subsection{SGL time delay}
For an SGL system, as the light from the source passes around a massive galaxy as the lens, it is split into multiple rays that take different paths and pass through different gravitational potentials resulting in a time delay to reach the observer. The measured time delay between two images ($\theta_{i}$ and $\theta_{j}$) is related to the time-delay distance $D_{\Delta \mathrm{t}}$ and the Fermat potential difference $\Delta \phi_{i, j}$ as

\begin{equation}
\Delta t_{i, j}=\frac{D_{\Delta \mathrm{t}}}{c} \Delta \phi_{i, j}.
\label{Eq:Eq8}
\end{equation}
Here, the time-delay distance is the combination of the three angular diameter distances \citep{Refsdal:1964nw,Suyu:2009by}
\begin{equation}
D_{\Delta \mathrm{t}}=\frac{c \Delta t_{i, j}}{\Delta \phi_{i, j}}=\frac{c}{H_{0}} \frac{d_{l} d_{s}}{d_{l s}}.
\label{Eq:Eq9}
\end{equation}

The mass-sheet transformation (MST) is an important factor to be considered in lens modelling (for a detailed discussion see \citep{Schneider:2013sxa,Chen:2019ejq}). It can transform a projected mass distribution of lens galaxy $\kappa(\theta)$ into infinite sets of $\kappa_{\lambda}(\theta)$,
\begin{equation}
\kappa_{\lambda}({\theta})=\lambda \kappa({\theta})+(1-\lambda),
\label{Eq:MST}
\end{equation}
with the same dimensionless observables. In other words, with the observed image positions, image shapes, magnification ratios, flux ratios, etc, it is not possible to distinguish the original $\kappa({\theta})$ from $\kappa_{\lambda}({\theta})$ in Equation (\ref{Eq:MST}), so-called mass-sheet degeneracy (MSD). The MSD is one of the main sources of uncertainty in time-delay distance measurement \citep{Suyu:2013kha,2013ApJ...766...70S}. An effective method to break the MSD is combining lensing measurements with stellar kinematics measurements of the lens galaxy \citep{Schneider:2013sxa,Chen:2019ejq}.

Besides, the additional mass along the LOS between the observer and the source should also be considered. This effect could effectively be approximated by a constant external convergence term, $\kappa_{\mathrm{ext}}$, which can be estimated by the constraint on the mass along the LOS to the lens from observational data \citep{Wong:2019kwg}. Thus, the true time-delay distance could be written as 
\begin{equation}
D_{\Delta t}= \frac{D_{\Delta t}^{\mathrm{model}}}{1-\kappa_{\mathrm{ext}}},
\end{equation}
where $D_{\Delta t}^{\mathrm{model}}$ is the time-delay distance inferred from the lens model and measured time delays \citep{Wong:2019kwg}.

At present, the state-of-the-art time delays measurements come from COSMOGRAIL (COSmological MOnitoring of GRAvItational Lenses) programme \citep{Bonvin:2015jia,Eulaers:2013iha,DES:2017uif,Millon:2020xab,Vuissoz:2008qa,Tewes:2012zz}. Based on the high-quality lensed quasars obtained via optical monitoring by COSMOGRAIL, the H0LiCOW ($H_0$ Lenses in COSMOGRAIL's Wellspring) program presented a series of precise measurements of $H_0$ \citep{Suyu:2016qxx,Wong:2019kwg,Bonvin:2016crt,Birrer:2018vtm,Chen:2019ejq,Rusu:2019xrq,Wong:2016dpo}. The SGL time-delay observational sample we use include six lensing systems released by the H0LiCOW collaboration: B1608+656 \citep{Suyu:2009by, Jee:2019hah}, RXJ1131-1231 \citep{Chen:2019ejq, Suyu:2012aa}, HE 0435-1223 \citep{Wong:2016dpo, Chen:2019ejq}, SDSS 1206+4332 \citep{Birrer:2018vtm}, WFI2033-4723 \citep{Rusu:2019xrq} and PG 1115+080 \citep{Chen:2019ejq}. Among these SGL systems, the range of source redshift is 0.654 to 1.789. The detailed analysis and relevant parameters for these six SGL systems include redshifts of sources and lenses, along with the posterior distributions of time delay distance $D_{\Delta \mathrm{t}}$ in the form of Monte Carlo Markov chains (MCMC), which are all summarized in \cite{Wong:2019kwg}. It should be noted that the likelihood function $\mathcal{L}_{D_{\Delta \mathrm{t}}}$ for each SGL system is calculated by using kernel density estimation.

\section{Results and Discussions} \label{sec:results}
We use the \texttt{emcee} Python module \citep{Foreman-Mackey:2012any} based on MCMC to constrain the Hubble constant and the spacial curvature, along with the polynomial coefficients ($a_1$, $a_2$). The final likelihood $\mathcal{L}$ using in our analysis is the combination of $\mathcal{L}_{\mathrm{GW}}$ and $\mathcal{L}_{D_{\Delta_{t}}}$, which can be written as 

\begin{equation}
\ln \mathcal{L}=\ln \left(\mathcal{L}_{\mathrm{GW}}\right)+\ln \left(\mathcal{L}_{D_{\Delta t}}\right).
\label{Eq:Eq11}
\end{equation}

\begin{table*}[htbp]
\begin{center}{
\caption{Constraints on $H_0$, $\Omega_K$, and the coefficients of third-order polynomial ($a_1$, $a_2$) with 1$\sigma$ confidence level from SGLTD and GW data in the framework of distance sum rule.  \label{table:1}}
\begin{tabular}{c|cccc} \hline\hline
Data Set &$H_0~({\rm km~s^{-1}~Mpc^{-1}})$  &$\Omega_K$  &$a_1$  &$a_2$ \\
  \hline
\multirow{3}{*}{6 observed SGLTD + 1000 simulated GW}& $73.69\pm{0.36}$  &$0.076_{-0.087}^{+0.068}$ & $-0.274\pm{0.010}$  &$0.035\pm{0.005}$ \\
 &$73.66\pm{0.36}$  &$0~({\rm fixed})$ &$-0.273\pm{0.010}$  &$0.035\pm{0.005}$ \\
 &$74.03~({\rm fixed})$  &$0.082_{-0.087}^{+0.072}$ &$-0.266\pm{0.006}$  &$0.032\pm{0.003}$ \\
    \hline\\
55 simulated SGLTD + 1000 simulated GW & $73.65\pm0.35$ & $0.008\pm 0.048$ & $-0.274\pm0.010$ & $0.0357\pm 0.0045$\\
\hline
\hline
 \end{tabular}
 }
\end{center}
\end{table*}

\begin{figure*}[htbp]
\centering
\includegraphics[scale=0.5]{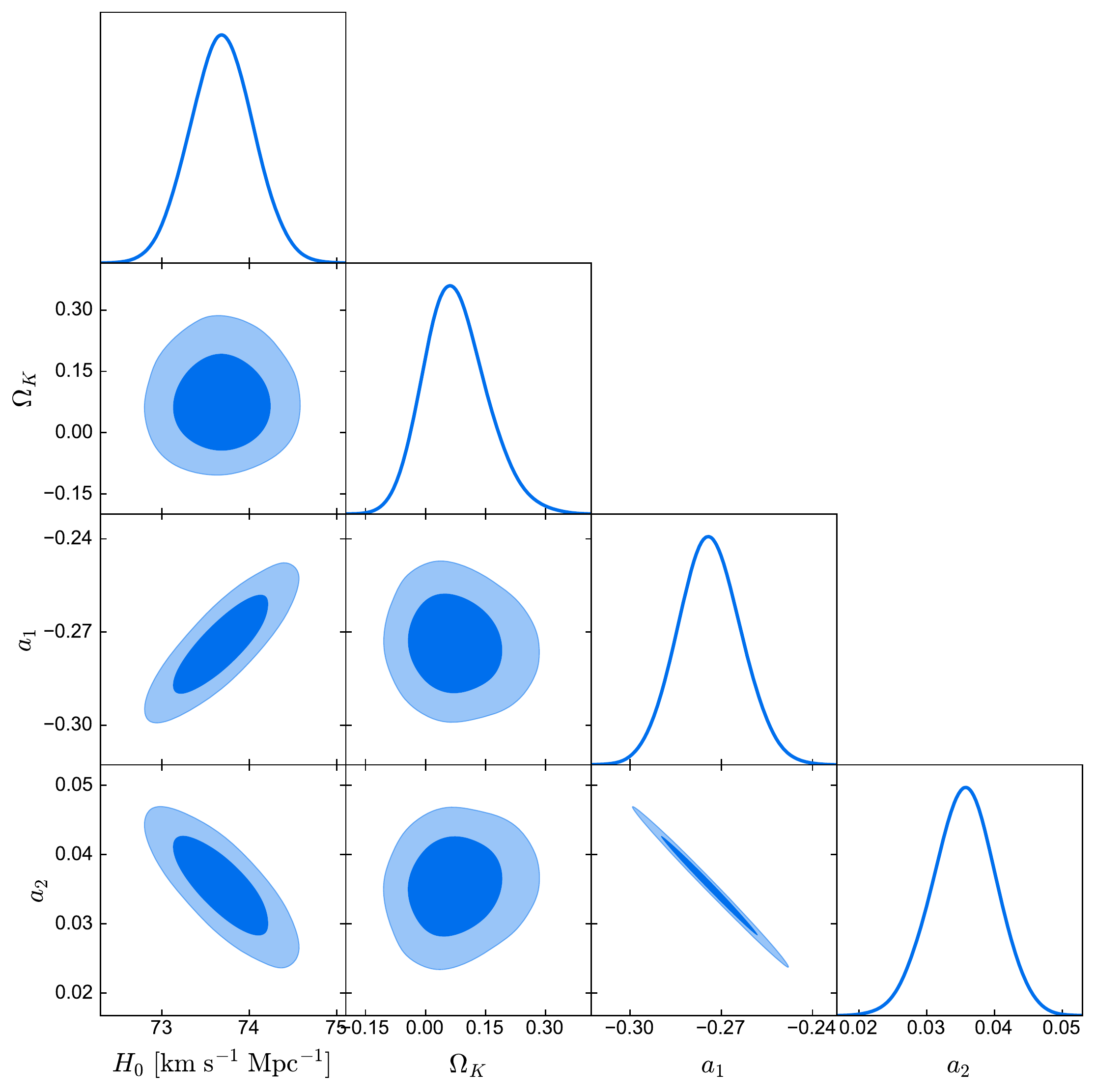}
\caption{1D and 2D marginalized probability distributions with the 1$\sigma$ and 2$\sigma$ contours for $H_0$, $\Omega_K$, and the coefficients of third-order polynomial ($a_1$, $a_2$) constrained by using 6 observed SGLTD and 1000 simulated GW data.}
\label{fig:fig2}
\end{figure*}

\subsection{Constraints on $H_0$ and $\Omega_K$ with simulated GW and 6 observed SGL} 
Based on the distance sum rule, we constrain $H_0$ and $\Omega_K$ from the combination of 6 observed time delay measurements of SGL systems and 1000 simulated GW standard siren data observed from the ET in the future. The constraint results with 1$\sigma$ confidence level for $H_0$, $\Omega_K$ and the polynomial coefficients ($a_1$ and $a_2$), are listed in Table \ref{table:1}, and the corresponding 1D and 2D marginalized probability distributions are shown in Figure \ref{fig:fig2}. It is worth mentioning that, the constraint on $H_0$ we obtain is $H_0=73.69\pm0.36 \mpc$ with a 0.5 \% uncertainty, which is much better than the \textit{Planck} 2018 results $H_0=67.4\pm0.5 \mpc$ with a 0.7 \% uncertainty \citep{Planck:2018vyg}. This constraint precision has met the requirements of precision cosmology, which shows that our method can effectively explore the Hubble tension problem.

For the cosmic curvature parameter $\Omega_K$, the best-fit value with 68.3 \% confidence level we obtain is $\Omega_K=0.076_{-0.087}^{+0.068}$, which shows no significant deviation from zero. Although the GW data are simulated based on a flat universe, the 6 SGLTD data are observed, so the constraint result of $\Omega_K$ more or less could be used as support for a flat universe. The interesting thing is that the degeneracy between $H_0$ and $\Omega_K$ is very weak as shown in the 2D marginalized contours on $H_0-\Omega_K$ plane in Figure \ref{fig:fig2}, which is different from the results obtained in previous related works with a positive correlation \citep{Qi:2020rmm,Collett:2019hrr,Wei:2020suh}. A weak degeneracy between two parameters means that the constraint precision of one parameter is weakly affected when another parameter is determined, which also could be seen clearly in Table \ref{table:1}. Assuming a flat universe ($\Omega_K=0$), we get $H_0=73.66\pm{0.36}$ that is almost the same as the results in the case of $\Omega_K$ as a free parameter. With the prior of $H_0=74.03~\mpc$, we get $\Omega_K=0.082_{-0.087}^{+0.072}$, as expected, which is also the same as one obtained without the prior.

\begin{figure}[htbp]
\centering
\includegraphics[scale=0.6]{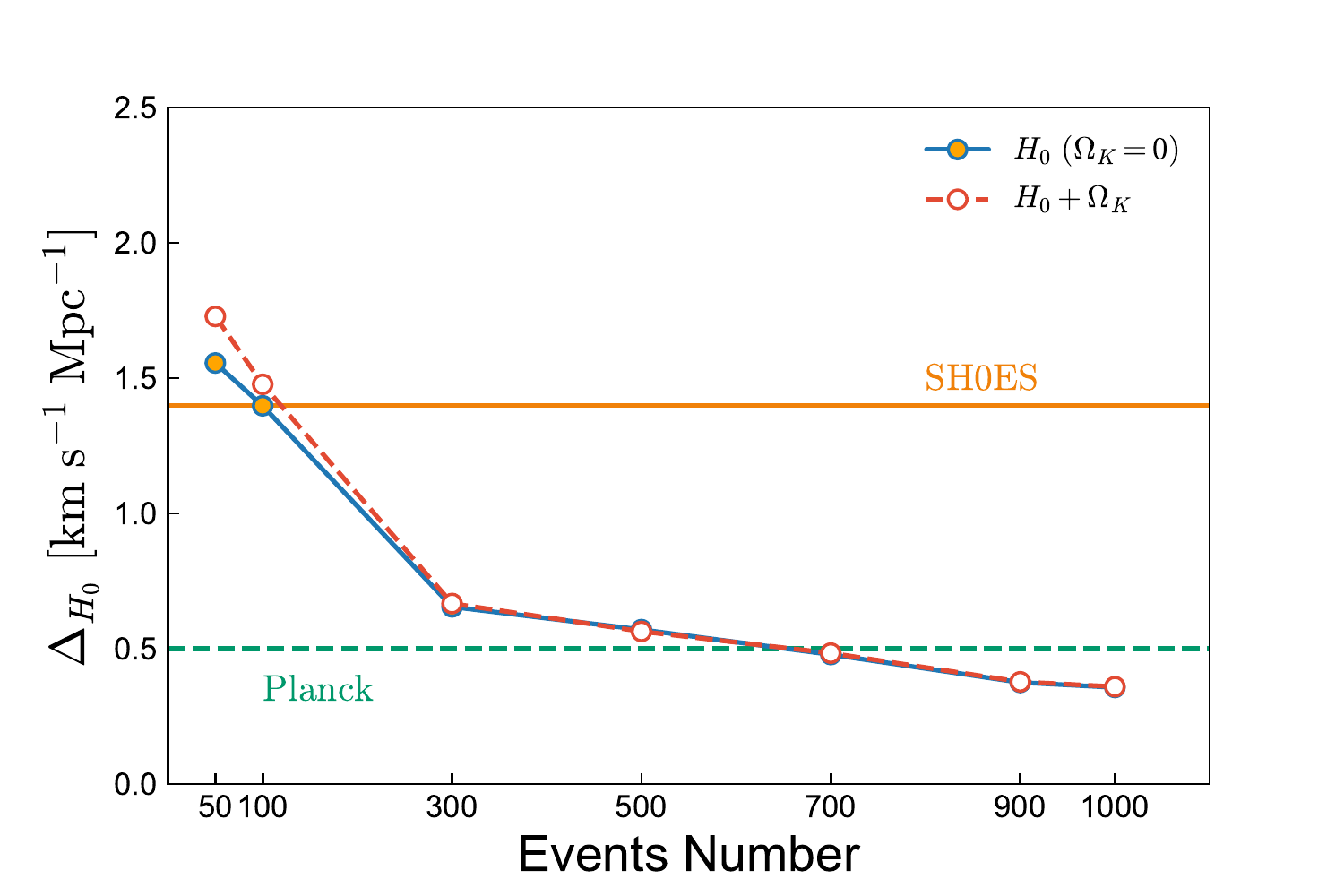}
\caption{The constraint precision of $H_0$, $\Delta_{H_0}$, for different numbers of GW events. The blue and red polylines are the results of taking the prior $\Omega_K=0$ and $\Omega_K$ as a free parameter together, respectively. The light yellow line and green dotted line represent the precision of $H_0$ measured by SH0ES and \textit{Planck} respectively.
}
\label{fig:fig3}
\end{figure}

On the other hand, to further investigate the constraint ability of this model-independent method, we constrain $H_0$ and $\Omega_K$ in the cases that the detected number of GW events reaches 50, 100, 300, 500, 700, 900, 1000, respectively. Figure \ref{fig:fig3} shows the improvement in the constraint precision of $H_0$, $\Delta_{H_0}$, as the detected GW events accumulate. Obviously, with the increase of detected GW events, the constraints on $H_0$ improve significantly. The light yellow line and green dotted line represent the precision of $H_0$ measured by SH0ES and \textit{Planck} respectively. As we can see, the constraint precision for $H_0$ at about 100 GW events is comparable to the measurement from SH0ES collaboration \citep{Riess:2019cxk}. As the number of GW events increases, constraint precision at 700 GW events could exceed that of the \textit{Planck} 2018 results \citep{Planck:2018vyg}. With and without the prior of $\Omega_K=0$, the constraint precision of $H_0$ has a slight difference at a small number of GW events, $N=50$ for example, which means there is a degeneracy between $H_0$ and $\Omega_K$. This seem s to lead to the conclusion that degeneracy will be affected by the number of detection events in this method.

\begin{figure}[htbp]
\centering
\includegraphics[scale=0.6]{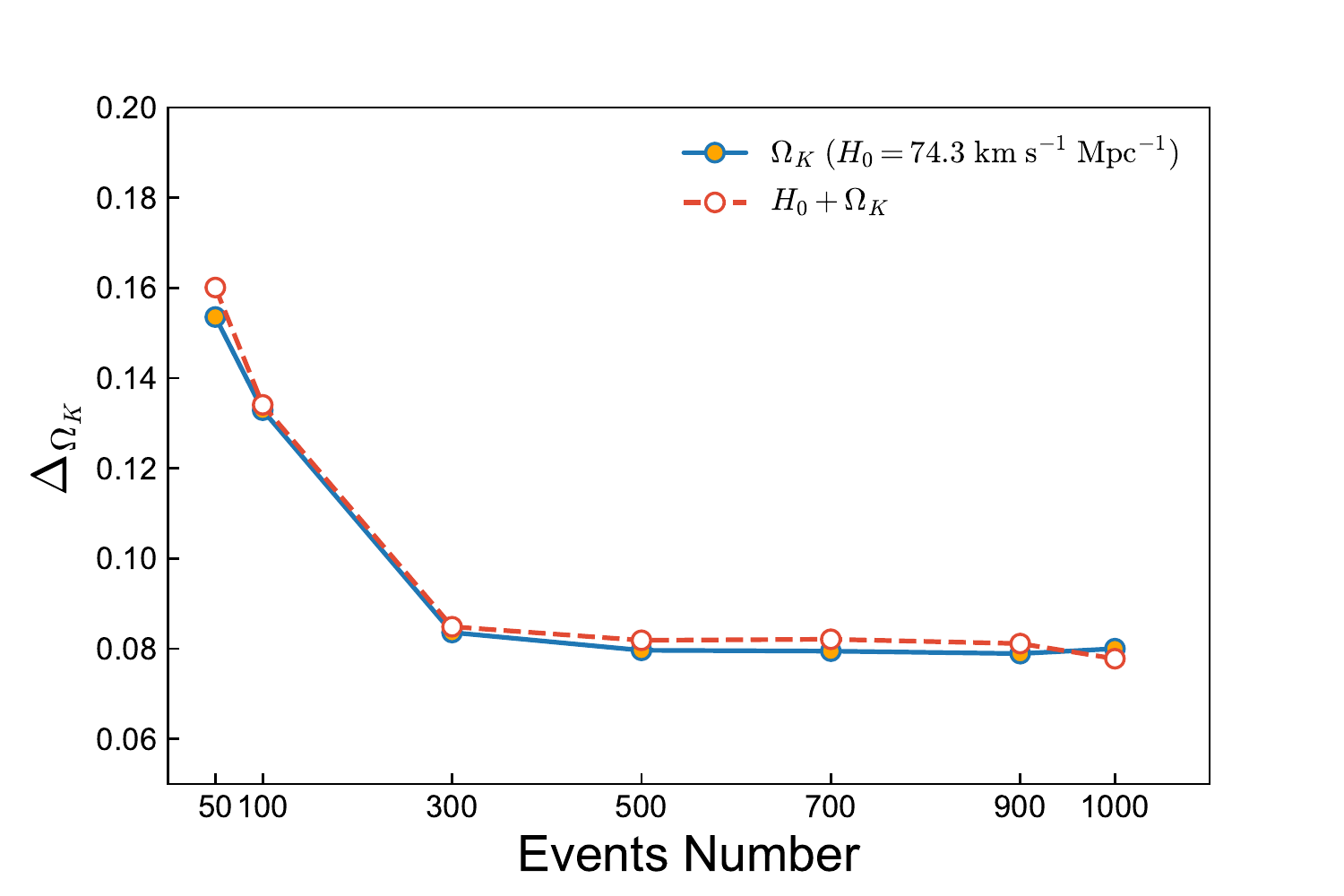}
\caption{The constraint precision of $\Omega_K$, $\Delta_{\Omega_K}$, for different numbers of GW events. The blue and red polylines are the results of taking the prior $H_0= 74.03~\mpc$ and $H_0$ as a free parameter together, respectively.}
\label{fig:fig4}
\end{figure}

In Figure \ref{fig:fig4}, we present the improvement in the constraint precision of $\Omega_K$, $\Delta_{\Omega_K}$, as the number of GW events increase. We find that the constraints of $\Omega_K$ improve significantly with the increase of GW events from 50 to 300, but it is almost no longer improved after 300 GW events. Such a trend also exists in the constraints on $H_0$ as shown in Figure \ref{fig:fig3}. This reminds us that we more need to improve the precision of the measurement rather than increase the number of GW events after about 300 detected GW events. In addition, with and without the prior of $H_0= 74.03~\mpc$, the constraint precision of $\Omega_K$ is almost the same.

\subsection{Constraints on $H_0$ and $\Omega_K$ with simulated GW and simulated SGLTD} 

\begin{figure}[htbp]
\centering
\includegraphics[scale=0.6]{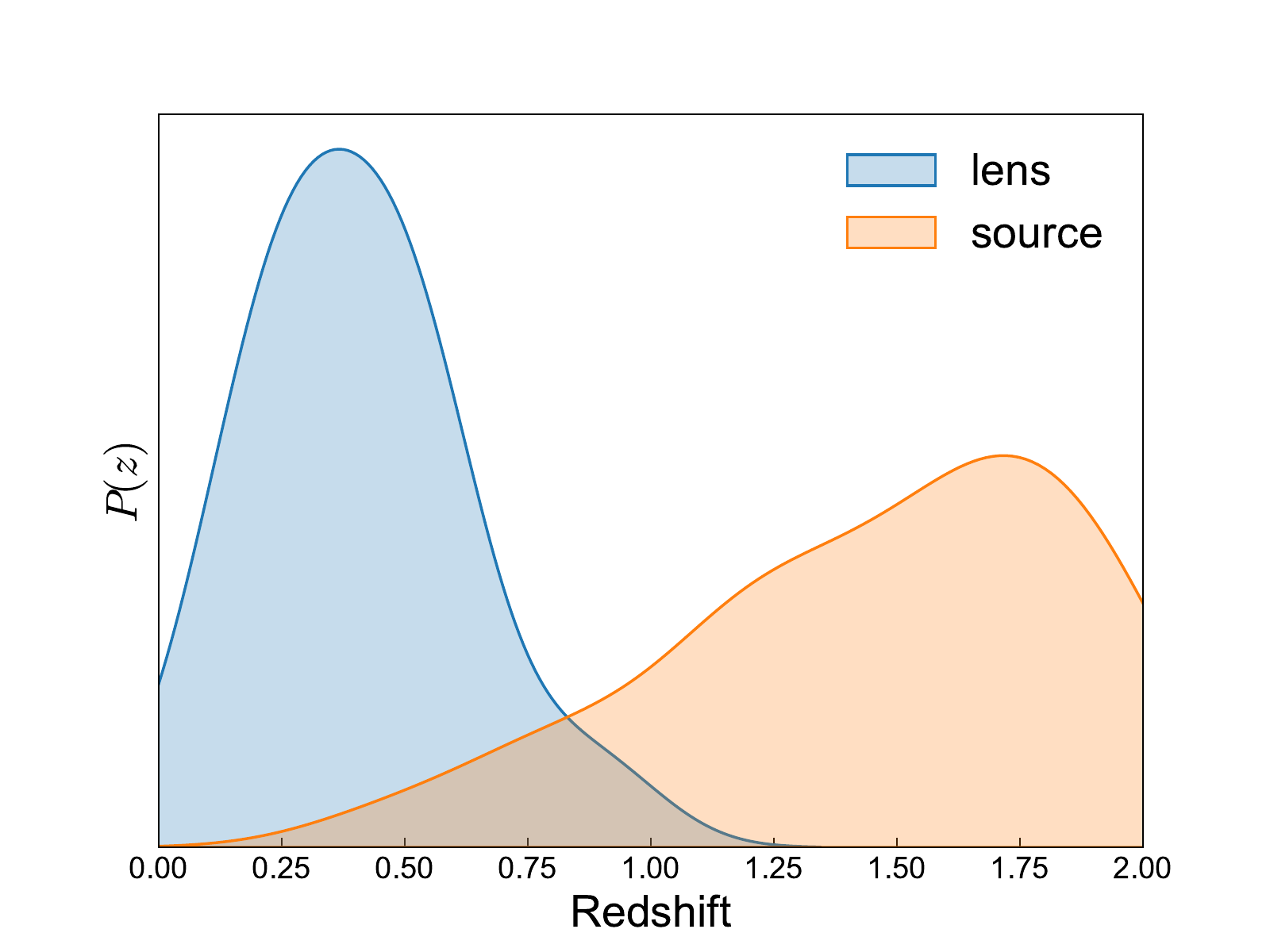}
\caption{The redshift distribution of the sources and lenses for the simulated SGLTD data. }
\label{fig:zl_zs}
\end{figure}

At the same time as the construction and observation of the third-generation gravitational wave detector, the ongoing and future massive surveys with high-quality imaging and spectroscopy such as \textit{Euclid} Survey and LSST will provide a large sample of well-measured SGLTD observations. For instance, based on the forecasting prescriptions in the previous work \citep{Oguri:2010ns}, LSST will find more than 8000 lensed quasars, of which about 400 data will have well-measured time delay. However, the measurement for time-delay distance we used in this paper also requires the central velocity dispersion of the lens galaxies and the LOS measurements that are necessary to break the mass-sheet degeneracy and to estimate the external convergence. Started with the mock catalogue of lensed QSO expected for LSST from \citet{Oguri:2010ns}, by selecting with strict criteria \citep{Jee:2015yra,Wen:2019yem}, we randomly generate a mock sample containing 55 SGL systems with time delay distance and angular diameter distances measurements. Figure \ref{fig:zl_zs} shows the redshift distribution of the sources and lenses for these mock SGL systems. Strictly speaking, the different redshift distributions of SGL systems will affect the following final results, because the system at lower redshift is brighter and would yield a good delay. However, our simulation is generated randomly by adopting the realistic distributions for the lens and source properties without loss of generality. According to the realistic current lensed quasar constraints \citep{Suyu:2020opl,Chen:2019ejq,Tewes:2012zz,Vuissoz:2008qa,Bonvin:2016crt,Suyu:2016qxx,Millon:2020ugy,Millon:2020xab,DES:2017uif}, we conservatively adopt 5\% for the time-delay uncertainties, 3\% for the lens mass modelling uncertainties, and 3\% for the lens environment uncertainties. Adding these in quadrature, we set 6.6\% uncertainty to $D_{\Delta t}$. We also adopt the $\Lambda$CDM model with cosmological parameters $\Omega_{m} = 0.315$, $H_{0} = 74.03 \mpc$ as the fiducial model in the SGLTD simulation.

\begin{figure}[htbp]
\centering
\includegraphics[scale=0.4]{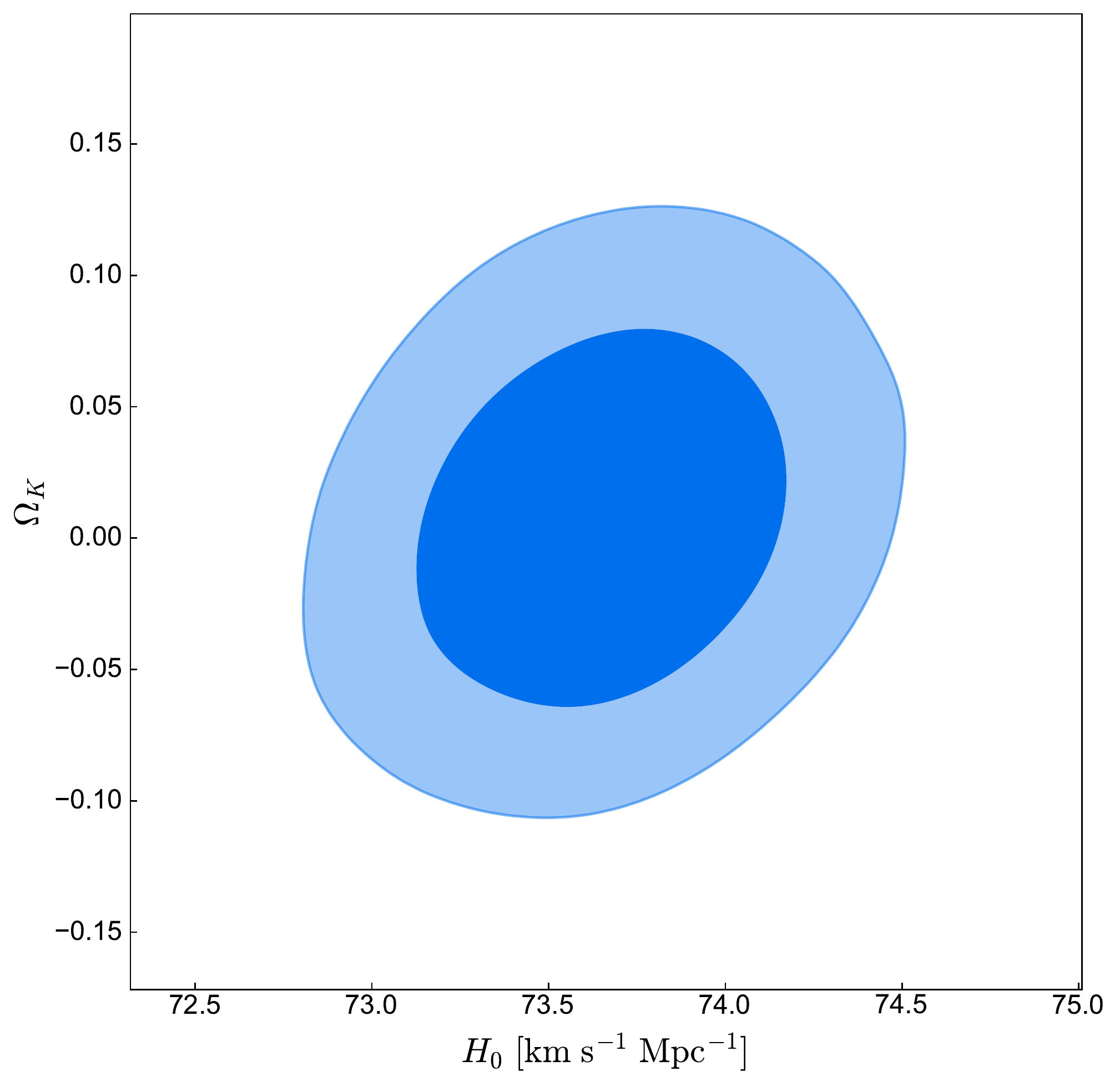}
\caption{1D and 2D marginalized probability distributions at 68.3\% and 95.4\% confidence level for $H_0$ and $\Omega_K$ constrained by using 55 simulated SGLTD and 1000 simulated GW data.}
\label{fig:fig5}
\end{figure}

In Figure \ref{fig:fig5} and Table \ref{table:1}, we show the constraint results of $H_0$ and $\Omega_K$ obtained by the combination of 55 simulated SGLTD and 1000 simulated GW. We find that there is a slight positive correlation between $H_0$ and $\Omega_K $ inconsistent with previous related works \citep{Qi:2020rmm,Collett:2019hrr,Wei:2020suh}, which seems to confirm our conclusion mentioned above that the number of observed samples affect the degeneracy of those two parameters in this method. For the constraint precision of parameters, it is seen that the constraint on the Hubble constant, $H_0=73.65\pm0.35 \mpc$, is improved barely compared with the results from 6 observed SGLTD. But for the cosmic curvature parameter, we obtain $\Omega_K=0.008\pm0.048$ from the simulated SGLTD data, which is improved by about 50\% compared with the results from actually observed SGLTD. For the best-fit value of $\Omega_k$, we find it shifts obviously from 0.076 obtained from actually observed SGLTD to 0.008 here, although both of them are in agreement well with the flat universe within their 1 $\sigma$ confidence level. One of the possible reasons for its best-fit value changing we suspect is the influence of the lens mass distribution modeling. As mentioned above, for the measurements of SGLTD, an important and difficult factor is that the mass distribution of the lens galaxy needs to be determined with sufficient accuracy. Due to the insufficient understanding of the mass distribution in the inner parts of galaxies, parametrized mass models are usually employed. The deviation between the model and the actual galaxy may lead to a bias in parameter estimation. As future massive surveys observe more and more SGL systems, a more accurate model for lens galaxies is expected to be obtained, which will be very helpful to the constraint on cosmological parameters. Combining this constrained result with Figure \ref{fig:fig4}, it can be seen that the constraint precision of $\Omega_K$ will not be improved with the increase of the GW number when the number of GW reaches a certain number, but the increase of SGLTD number will have a greater impact on it.

\section{Conclusion} \label{sec:conclusion}

With the improvement of the number and the precision of observations, the inconsistencies for some key cosmological parameters have arisen, $H_0$ and $\Omega_K$ for example, which in essence reflects that our understanding of the universe may be flawed under the framework of standard cosmological theory. At present, further confirming these inconsistencies in different ways is necessary. In particular, it is of importance to measure fundamental cosmological parameters in the late universe with some new and model-independent ways. In this paper, based on the distance sum rule, we present a new way to constrain $H_0$ and $\Omega_K$ simultaneously in the late universe with the combination of the GW standard siren data from the future observation of the ET and the SGLTD data.

Based on the currently 6 observed SGLTD, the constraint precision for $H_0$ given by the combined 100 GW standard siren events can be comparable with the measurement from SH0ES collaboration. As the number of GW events increases to 700, the constraint precision of $H_0$ can exceed that of the \textit{Planck} 2018 results. When the number of GW standard siren events increases to 1000 as the conservative estimation of ET in ten-year observation, the constraint on $H_0$ could be improved to 0.5\% uncertainty. For the constraint on $\Omega_K$, we obtain $\Omega_K=0.076^{+0.068}_{-0.087}$ from 6 observed SGLTD+1000 simulated GW. Considering the influence of different numbers of GW events on the constraint precision of $\Omega_K$, we find that the constraints on $\Omega_K$ improve significantly with the increase of GW events from 50 to 300, but it is almost no longer improved after 300 GW events. Such a trend also exists in the constraints on $H_0$. This indicates that after the GW events number reaches about 300, the systematic errors will be dominant for the constraint precision of parameters rather than statistical errors.

On the other hand, at the same time as ET construction and observation, the LSST with wide field-of-view and high-quality imaging will provide a large sample of well-measured SGLTD observations. Based on LSST, we conservatively simulate 55 SGL systems with 6.6\% uncertainty for time-delay distance to investigate the potential of this cosmological model-independent constraint on $H_0$ and $\Omega_K$. Compared with the results from the combination of 6 observed SGLTD and 1000 simulated GW, we find that the constraint of $H_0$ from the combination of 55 simulated SGLTD and 1000 simulated GW is improved barely, while the constraint on $\Omega_K$ is improved about 50\%, i.e., $\Omega_K=0.008\pm0.048$. We conclude that to further improve the constraint of $\Omega_K$ depends more on the increase of the SGLTD observations than on the GW standard siren events when the number of GW reaches a certain number.

\section*{Acknowledgments}
This work was supported by the National Natural Science Foundation of China (Grants Nos. 11975072, 11835009, and 11875102), the Liaoning Revitalization Talents Program (Grant No. XLYC1905011), the National 111 Project of China (Grant No. B16009), and the Fundamental Research Funds for the Central Universities (Grant Nos. N2105014). We acknowledge the science research grants from the China Manned Space Project with NO.CMS-CSST-2021-B01.
\bibliography{h0_Ok}{}
\bibliographystyle{aasjournal}

\end{document}